\documentclass[preprint,aps,prl]{revtex4-1}
\usepackage{graphicx}
\usepackage{dcolumn}
\usepackage{bm}
\usepackage{amssymb}
\usepackage{color}
\usepackage{caption}
\usepackage{epstopdf}

\begin{document}
\title{Composition and Resolution Dependence of Effective Coarse-Graining Potentials in Multi-Resolution Simulations}
\author{M. Dinpajooh, M. G. Guenza\footnote{Corresponding author. Electronic mail: mguenza@uoregon.edu}}
\affiliation{Department of Chemistry and Biochemistry, and Institute of Theoretical Science, University of Oregon, Eugene, Oregon 97403}

\begin{abstract}
Given that the physical properties of polymeric liquids extend on a wide range of lengthscales, 
it is computationally convenient to represent them by coarse-grained (CG) descriptions
at various granularities to investigate local and global properties simultaneously.
This paper addresses how modeling the same system with mixed resolutions affects the consistency of the structural and thermodynamic properties, and shows that it is possible to formally derive interacting potentials that ensure consistency of the relevant physical properties in the mixed resolution region with the corresponding atomistic resolution simulations.
The composition, temperature, and density dependences of such Mixed Resolution Potentials (MRPs) are investigated.
In the limit of long polymer chains, where Markovian statistics is obeyed, the MRPs are analytically solved and 
decay with characteristic scaling exponents that depend on the mixture composition and CG resolution of the two components. 
Adopting MRPs simplifies the structure of multi-resolution simulations while quantitatively producing the structural and thermodynamical properties of the related atomistic systems such as radial distribution function and pressure.

\end{abstract}

\maketitle

In molecular liquids, understanding important phenomena such as phase transitions, nucleation, and crystal growth 
involves analyzing properties related on multiple scales in a complicated way.
\cite{Palmer2014,Radu2017,Fritz2011,Schweizer1990,Frenkel2002}
Simultaneous access to different lengthscale resolutions, by modulating the granularity in the modeling of the molecules, has the advantage of being computationally efficient,\cite{Dinpajooh2018} while retaining the desired information on all the lengthscales of interest. 
In this spirit, a number of multi-resolution approaches have been developed, both at the quantum\cite{Warshel1976,Warshel2003}, and at the 
classical levels.\cite{DelleSite2007,Potestio2013,Potestio2013b}
At the classical level, the most successful method is the Adaptive Resolution Simulations (AdResS) scheme, which involves density fluctuations and diffusion of molecules from
the high resolution, often atomistic, region to a low-resolution coarse-grained  region (and vice versa).
However, molecules represented at different granularity interact through effective CG potentials that depend on both molecular and physical parameters. When in a simulation, regions represented at different resolutions are brought together, inaccurate CG potentials produce gradient in pressure and chemical potential, with the emergence of unwanted spurious forces. In AdResS, this problem has been mitigated by constructing an intermediate region, the so-called hybrid region,  where molecules at different resolutions are mixed, and properties are gradually changed by the use of an empirical  'switch function'. Introducing the hybrid region leads to loss of conservation of either the energy or the momentum.

The question, thus, remains if it is possible at all, in a generic multiscale resolution scheme, to construct a mixed resolution region, where molecules represented by different levels of coarse-graining are thermodynamically and structurally compatible; and if this mixed region can maintain consistencies of structural and thermodynamic properties with the related atomistic description. Thermodynamics and structural properties of the multi-resolution approach should be independent of the CG representation, while correctly maintaining the properties of the underlying atomistic system. 

This study addresses these questions by building a mixed resolution region where molecules with different CG resolutions are shown to have thermodynamic and structural properties consistent with the atomistic descriptions. The study also answers the questions of how in a multi-resolution scheme the properties of the effective CG potentials, acting between molecules with equal and with different CG resolutions, are affected by the granularity of the models 
selected, and how the physical parameters characterizing the mixtures, including volume fraction, density, and 
temperature, affect the Mixed Resolution Potentials (MRPs). We take advantage of the fact that mixtures of multi-resolution polymeric liquids afford an analytical solution of the MRPs, 
because at low enough resolution the distributions of the CG sites along a long polymeric chain obey Markovian statistics.\cite{Clark2012,Clark2013}   Starting from the Integral Equation theory of Coarse-Graining (IECG),\cite{Clark2012,Dinpajooh2018,Guenza2018a} which combines  the PRISM approach at the atomistic level\cite{Schweizer1987} with our CG model, we extend the theory here to mixtures of identical polymer chains, but represented at two different CG resolutions. The granularity of the mixed resolution model adopted is low because it is only for these low resolution CG mixtures that it is possible to solve analytically the problem at hand. The method is demonstrated in the canonical ensemble, but other ensembles are possible. 

We study a binary blend mixture of $\alpha$ and $\beta$ homopolymers with degrees of polymerization of $N_\alpha$ and $N_\beta$.
The polymer volume fraction for polymers with the same segmental volume is $\phi = n_\alpha N_\alpha / (n_\alpha N_\alpha + n_\beta N_\beta)$, where $n_\alpha$ and $n_\beta$ are the number of
chains of type $\alpha$ and $\beta$ in the mixture, respectively.\cite{Schweizer1990}
In addition, the monomer density for a given species of $\alpha$ is defined as $\rho_{m \alpha}= \rho_{m} \phi_\alpha$,
where $\rho_m = (n_\alpha N_\alpha + n_\beta N_\beta)/V$, and $V$ is the volume of the system.
Similarly, the chain density for a given species of $\alpha$ is defined as $\rho_{c \alpha}= \rho_{m\alpha}/N_\alpha$.

\begin{figure}[htb]
\includegraphics[width=0.8\columnwidth]{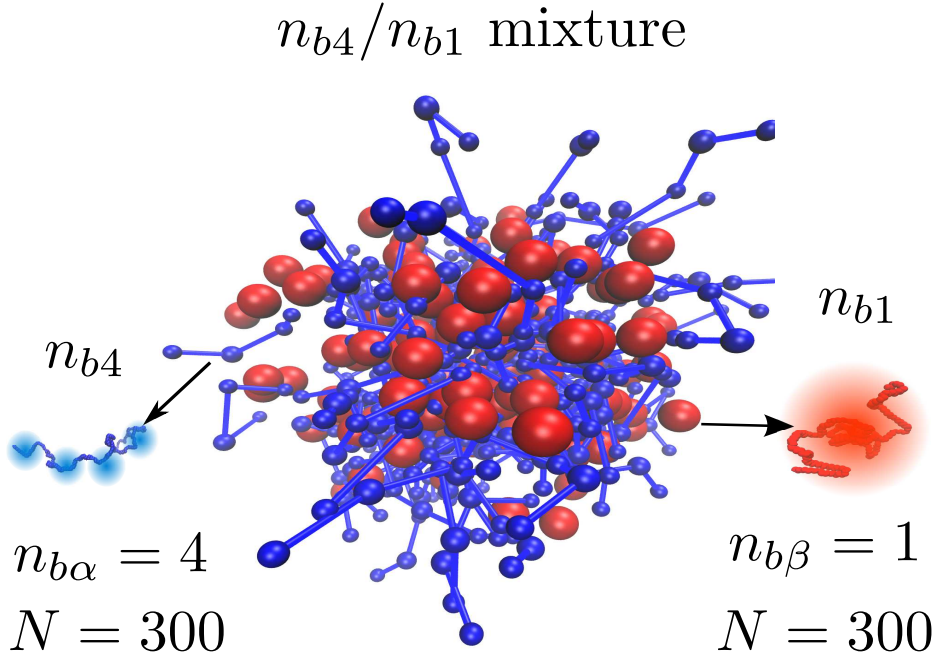}
\caption{ A snapshot from the IECG MD simulation trajectory of a binary mixture of 
polymer chains of linear polyethylene with one degree of polymerization and identical chemical structure, but different CG resolutions. The two components have the same
mean-square end-to-end distance and $c_0$ values, while the first component is represented by a chain of $4$ CG sites ($n_{b4}$) and the second component by a single CG site ($n_{b1}$).}
\label{iecgmix}
\end{figure}

In the IECG approach, each polymer is coarse-grained into a number of CG sites (also called blobs), $n_{b\alpha}$, and the density of blobs
for a given species, $\rho_{b\alpha}$, is related to $\rho_{c\alpha}$ via $\rho_{b\alpha} = \rho_{c\alpha} n_{b\alpha}$.
Figure \ref{iecgmix} shows a snapshot of the IECG simulations for a mixed resolution mixture with $n_{b\alpha}=4$ and $n_{b\beta}=1$. In a shorthand notation this mixture is reported as $n_{b4}/n_{b1}$, where $n_{b4}$ means that $n_{b\alpha}=4$.

The generalized Ornstein-Zernike (OZ) matrix relation\cite{Yatsenko2004} for this binary system
is expressed as a function of monomer sites and the CG (auxiliary) sites structural distributions in Fourier space as

\begin{equation}
\mathbf{\hat{H}}(k)=\mathbf{\hat{\Omega}}(k)\mathbf{\hat{C}}(k)\left[\mathbf{\hat{\Omega}}(k)+\mathbf{\hat{H}}(k)\right],
\label{MatrixOZ}
\end{equation}
where $\mathbf{\hat{H}}(k)$, $\mathbf{\hat{\Omega}}(k)$, and $\mathbf{\hat{C}}(k)$ are 
block matrices of rank 4 representing the total,
intramolecular, and direct correlation functions for the aforementioned binary system, respectively.\cite{Yatsenko2004}
Considering the intra and inter molecular correlations between monomers and CG units,  and between CG units, and 
approximating the monomer-monomer direct correlation functions by their large-scale $k\rightarrow 0$ behavior ($c_0$), 
the IECG equation for a mixture gives the total self and cross correlation functions between CG (blob) sites as

\begin{equation}
\hat{h}_{\alpha\alpha}^{bb}(k) = -\frac{ \hat{\Omega}_{\alpha}^{mb}(k)^2 \left( n_{b\alpha} \Gamma_{b\alpha}  + \delta \Gamma \hat{\Omega}_{\beta}^{mm}(k) \right) }
{ \rho_{c\alpha} \Lambda },
\label{hbbAA}
\end{equation}
where $\hat{h}_{\beta\beta}^{bb}(k)$ is obtained by interchanging $\alpha$ and $\beta$ labels in the formula for $\hat{h}_{\alpha\alpha}^{bb}(k)$ and
\begin{equation}
\hat{h}_{\alpha\beta}^{bb}(k) =- \left( \frac{n_{b\alpha} n_{b\beta}}{\rho_{c\alpha} \rho_{c\beta}}  \right)^{1/2}  \frac{ \Gamma_{b\alpha\beta} \hat{\Omega}_{\alpha}^{mb}(k) \hat{\Omega}_{\beta}^{mb}(k) }
{ \Lambda },
\label{hbbAB}
\end{equation}
where $\Gamma_{b\alpha} = -N_{b\alpha} \rho_{m\alpha} c_{0\alpha\alpha}$, $N_{b\alpha}=N_{\alpha}/n_{b\alpha}$ and
$\Gamma_{b\alpha\beta} = -\sqrt{N_{b\alpha} N_{b\beta} \rho_{m\alpha} \rho_{m\beta} } c_{0\alpha\beta}$ and 
$\Lambda$ is given by

\begin{equation}
\Lambda = 1 + n_{b\alpha} \Gamma_{b\alpha} \hat{\Omega}_{\alpha}^{mm}(k) + n_{b\beta} \Gamma_{b\beta} \hat{\Omega}_{\beta}^{mm}(k)
 + \delta \Gamma \hat{\Omega}_{\alpha}^{mm}(k) \hat{\Omega}_{\beta}^{mm}(k),
\label{Lambda2}
\end{equation}
where $\delta \Gamma = n_{b\alpha} n_{b\beta} \left( \Gamma_{b\alpha} \Gamma_{b\beta} - \Gamma_{b\alpha\beta} \Gamma_{b\alpha\beta} \right)$ 
and $\Omega_\alpha^{mb}$ and $\Omega_\alpha^{bb}$ are the intramolecular correlation functions 
 that obey Gaussian statistics.\cite{Doi1986}

The direct correlation functions between CG sites in the mixture are then determined by solving the $2\times2$ OZ matrix
equation for a mixture consisting of the blob sites only:

\begin{equation}
\mathbf{\hat{C}^{bb}}(k) = \left( \mathbf{\hat{\Omega}^{bb}}(k) \right)^{-1}  \mathbf{\hat{H}^{bb}}(k)
\left[  \left(  \mathbf{\hat{\Omega}^{bb}}(k) + \mathbf{\hat{H}^{bb}}(k)  \right)  \right]^{-1}.
\label{OZCbb}
\end{equation}

In the small $k$ limit, the analytical solution of the direct correlations may be obtained by 
factoring out $\Gamma_{b\alpha}$, $\Gamma_{b\beta}$, and $\Gamma_{b\alpha\beta}$ from $\hat{c}^{bb}_{\alpha\alpha}(k)$,
$\hat{c}^{bb}_{\beta\beta}(k)$, and $\hat{c}^{bb}_{\alpha\beta}(k)$, respectively, and
rescaling $k$ to $k \Gamma_{b\alpha}^{-1/4}$, $k \Gamma_{b\beta}^{-1/4}$, and $k \Gamma_{b\alpha\beta}^{-1/4}$, respectively.
This allows one to expand the direct correlations in $k$ space and, after taking the limit of long polymer chains, one can neglect the contribution of large wave vectors and derive the direct self and cross correlation functions, which are thus approximated to the leading order as 

\begin{equation}
\hat{c}_{\alpha\alpha}^{bb}(k) \approx - \frac{\Gamma_{b\alpha}  N_{b\alpha} }{\rho_{m} \phi_\alpha} \frac{45 \left(45 \gamma_{\alpha}^2  + k^4 \zeta^4 ( \gamma_{\alpha} \gamma_{\beta} -1 ) \right) }
{45^2 \gamma_{\alpha}^2  + 45 k^4 \gamma_{\alpha} \left(  \gamma_{\alpha}+ \gamma_{\beta} \zeta^4 \right)  + k^8 \zeta^4 \left(  \gamma_{\alpha} \gamma_{\beta} -1  \right) },
\label{cbbAAanly}
\end{equation}
and
\begin{equation}
\hat{c}_{\alpha\beta}^{bb}(k) \approx - \frac{\Gamma_{b\alpha\beta}}{\rho_m} \sqrt{\frac{ N_{b\alpha} N_{b\beta} }{\phi_{\alpha}  (1-\phi_{\alpha})}} \frac{45^2}
{45^2 + 45 k^4 \left( \gamma_{\alpha} + \gamma_{\beta} \zeta^4 \right)  + k^8 \zeta^4 \left(  \gamma_{\alpha} \gamma_{\beta} -1  \right) },
\label{cbbABanly}
\end{equation}
with
\begin{equation}
\gamma_\alpha = \left( \frac{ \phi_\alpha N_{b\alpha} }{ (1-\phi_\alpha) N_{b\beta}  } \right)^{1/2} \frac{  c_{0\alpha\alpha}  }{ c_{0\alpha\beta} },
\label{gammaA}
\end{equation}
and
\begin{equation}
\gamma_\beta = \left( \frac{ (1-\phi_\alpha) N_{b\beta} }{ \phi_\alpha N_{b\alpha}  } \right)^{1/2} \frac{  c_{0\beta\beta}  }{ c_{0\alpha\beta} }.
\label{gammaB}
\end{equation}
Note that $k$ is expressed in units of  $R_{gb\alpha}$ and $\zeta=R_{gb\beta}/R_{gb\alpha}$, where $R_{gb\alpha} = R_{g\alpha}/\sqrt{n_{b\alpha}}$.
Additionally, $\Gamma_{b\alpha}  = \gamma_\alpha \Gamma_{b\alpha\beta}$,  $\Gamma_{b\beta}  = \gamma_\beta \Gamma_{b\alpha\beta}$, and
$\Gamma_{b\alpha} \gamma_\beta = \Gamma_{\beta}  \gamma_\alpha$. 

In the limit of a polymer binary mixture where the two components are given by the same polymer represented at two different levels of coarse-graining resolution, the equations are further simplified,
considering that $N_\alpha = N_\beta$, $R_{g\alpha}=R_{g\beta}$, and $\gamma_\alpha \gamma_\beta = 1$. 
 At liquid density it is possible to derive an analytical solution of the MRPs for the components in the multi-resolution mixture by considering that at liquid densities the closure to the generalized OZ equation can be assumed to follow the Mean Spherical Approximation (MSA).\cite{Likos2001,Clark2013} 
Furthermore, considering that the polymer chains are long enough to be well represented by chains of CG sites that follow a Markovian space distribution, the formal solution of the potentials for the multi-resolution mixture is readily calculated. 
Note that numerical potentials can be derived in more general cases of variable polymer density, length, and chemical nature by applying the HyperNetted Chain (HNC) closure approximation, 
which is very accurate for systems interacting through soft CG potentials. 
Numerical data from Molecular Dynamics (MD) IECG simulations reported in this paper make use of the HNC potential solved numerically.

In units of $R_{gb\alpha}$ the MRPs between the components of the multi-resolution mixture are given as

\begin{equation}
U_{\alpha\alpha}^{bb}(r) = k_{\rm{B}}T \frac{\Gamma_{b\alpha} \eta_\alpha^3  N_{b\alpha} }{2 \pi R_{gb\alpha}^3 \rho_{m} \phi_\alpha} \frac{ \sin (\eta_{\alpha} \: r) e^{- \eta_{\alpha} \: r} }{\eta_{\alpha} \: r } \ ,
\label{ubbAAr}
\end{equation}
with
\begin{equation}
\eta_{\alpha}= \sqrt{\frac{3}{2}} 5^{1/4} \left( \frac{1}{ \Gamma_{b{\alpha}}  (1+ \frac{\zeta^4}{\gamma_{\alpha}^2}) } \right)^{1/4} \ ,
\label{QAanly}
\end{equation}

and
\begin{equation}
U_{\alpha\beta}^{bb}(r) = k_{\rm{B}}T \frac{\Gamma_{b\alpha\beta} \eta_{\alpha\beta}^3  (N_{b\alpha}N_{b\beta})^{1/2}  }{2 \pi R_{gb\alpha}^3 \rho_{m} (\phi_\alpha (1-\phi_\alpha))^{1/2} } \frac{ \sin (\eta_{\alpha\beta} \: r) e^{- \eta_{\alpha\beta} \: r} }{\eta_{\alpha\beta} \: r } \ ,
\label{ubbABr}
\end{equation}
with
\begin{equation}
\eta_{\alpha\beta}= \sqrt{\frac{3}{2}} 5^{1/4} \left( \frac{ 1 }{ \Gamma_{b\alpha\beta}  (\gamma_{\alpha} +\gamma_{\beta} \zeta^4) }  \right)^{1/4} \ .
\label{QABanly}
\end{equation}
These equations are appropriate approximations for the long-range repulsive component of 
the inter bead potentials (see Figure S3).
The range of the MRP in units of $R_{gb\alpha}$ is found to
scale as $(\Gamma_{b\alpha} (1+\zeta^4/\gamma_\alpha^2) )^{1/4} \propto \rho_m^{1/4} c_0^{1/4} N_{b\alpha}^{1/4} (\phi_\alpha+ (1-\phi_\alpha) n_{b\alpha}^3/n_{b\beta}^3))^{1/4}$, where the direct correlation function is density and temperature dependent.
For a given species $\alpha$, as the granularity of the CG description increases (fine-grained representation),
the function $\Gamma_{b\alpha}$ decreases and, together with the mixture composition, they
determine the decay of the repulsive part of the MRP. 

\begin{figure}[htb]
\includegraphics[width=\columnwidth]{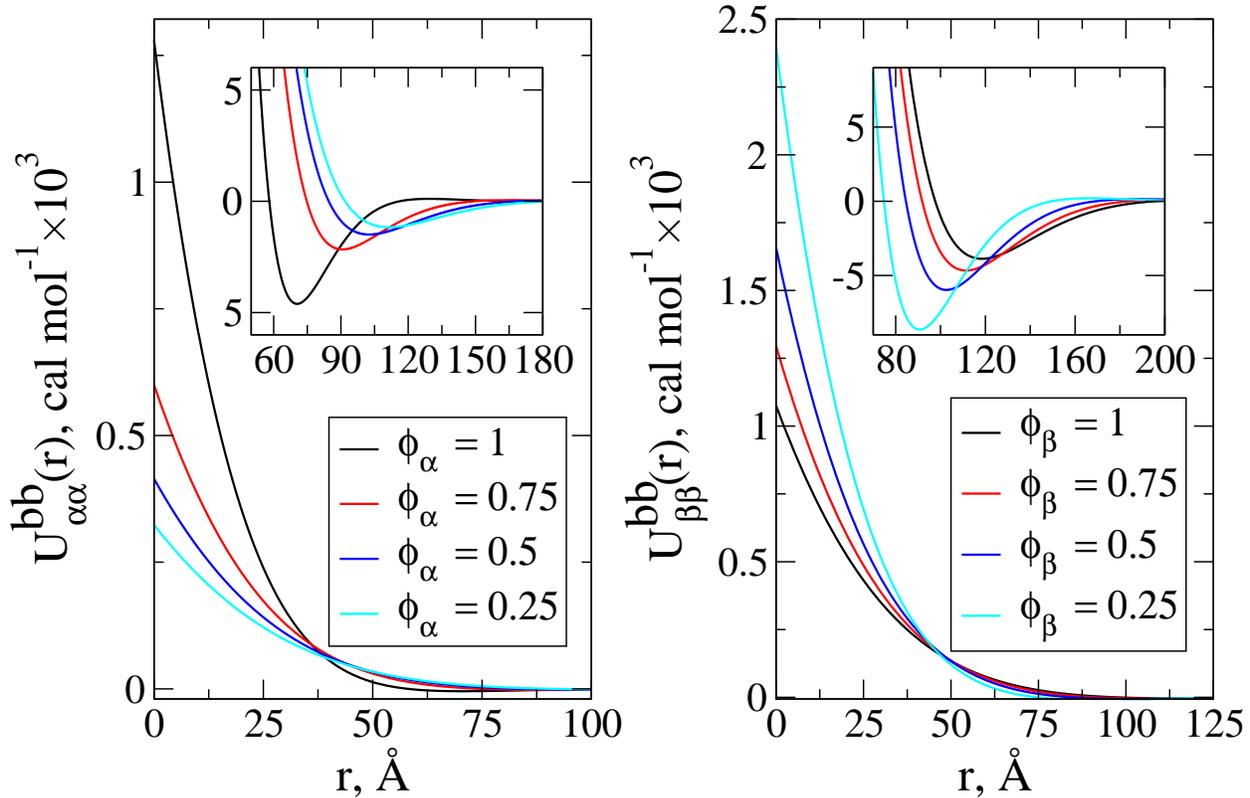}
\caption{The analytical effective coarse-graining (CG) potential for the mixture consisting of the polymer melts
with the degrees of polymerization $N=300$ represented by a mixture of four-blob chains ($\alpha$ species), and two-blob chains ($\beta$ species), at various compositions (volume fraction of species $\alpha$, $\phi_\alpha$) at $503$ K and 
a monomer density of $0.03296$ \AA$^{-3}$. In the inserts are details of the attractive part of the potentials.
Left: the effective CG potentials between CG sites consisting of $n_{b\alpha}=4$ blobs with $N_{b\alpha}=75$ monomers per blob.
Right: the effective CG potentials between CG sites consisting of $n_{b\beta}=2$ blobs with $N_{b\beta}=150$ monomers per blob.  
} 
\label{nb4nb2anly}
\end{figure}

The MRPs for a mixture consisting of chains represented by four CG sites ($n_{b4}$)
and chains represented by two CG sites ($n_{b2}$) are presented in Figure \ref{nb4nb2anly} for various mixture compositions. 
Here, a $c_{0}$ value of $-9.0$ \AA$^{-3}$ and a mean square end-to-end distance 
($\langle R^2 \rangle$) of $970$ \AA$^2$ have been used for all chains as an input to the IECG potential. 
These parameters are consistent with the chain dimension and liquid pressure of the atomistic simulation.
In this example, the polymer chains have degree-of-polymerization $N=300$, temperature $503$ K and a monomer density of $0.03296$ \AA$^{-3}$.
The range of the MRP between $n_{b4}$ species increases as the volume fraction of the $n_{b2}$ species increases; therefore, the range of MRP between $n_{b4}$ chains is more long-ranged in the mixture than in the pure polymer melts.
This is because in the mixtures, the interactions are mediated by the mixture containing $n_{b2}$ species, which are long-ranged: the many-body contributions to the pairwise potential between the $n_{b4}$ species
increase the range of their MRP.
On the other hand, as the composition of $n_{b2}$ species in the mixture increases,
the many-body effects for this species propagate to longer distances,
resulting in the increase of the range of MRP between the $n_{b4}$ species. 
It is worth mentioning that similar behavior is observed for the CG interactions between $n_{b4}$ 
and $n_{b2}$ species (results not shown).

\begin{figure}[htb]
\includegraphics[width=\columnwidth]{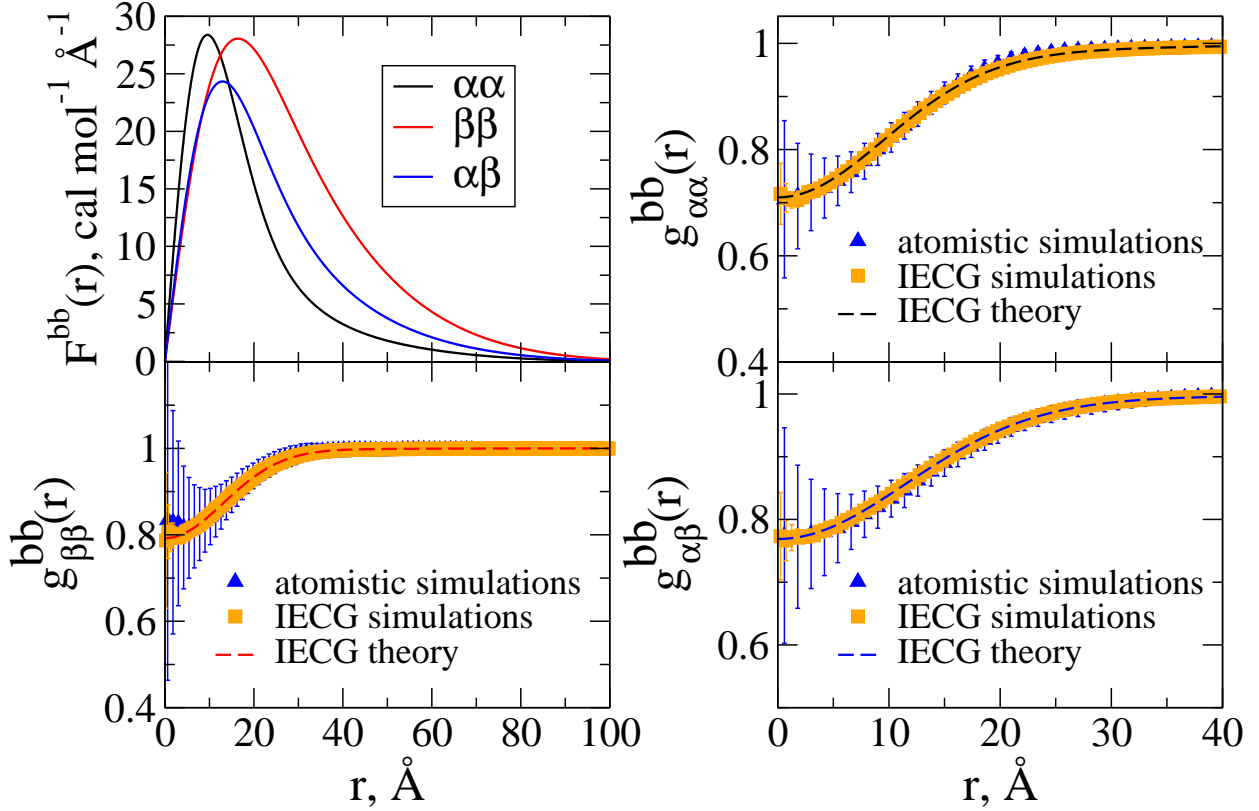}
\caption{ Comparison of the radial distribution functions (RDFs) for identical and different species present in the $n_{b4}/n_{b2}$ mixture at a volume
mole fraction of $\phi=0.5$. 
Left top panel: Effective CG forces between mixture components: note that chains are represented by $4$ CG sites ($\alpha$ species) and $2$ CG sites ($\beta$ species).
Top right panel: RDF between chains with $4$ CG sites. Bottom left panel: RDF between chains with $2$ CG sites. Bottom right: RDF between chains with $4$ and $2$ CG sites, respectively. 
The RDFs obtained by the IECG theory (solid lines) are within the statistical uncertainties of the RDFs from
the IECG simulations (orange squares),
and the ones from the atomistic simulations (blue triangles).}
\label{nb4nb2}
\end{figure}

Making use of the numerical HNC CG potential, one can directly compare the properties of the CG simulations to the atomistic ones, from which the starting parameters of  $c_0$ and polymer square end-to-end distance, $R^2$, were extracted.
The convergence of the related properties in the atomistic simulations are presented 
in the supplementary materials (SM).
Figure \ref{nb4nb2} shows the effective CG force between various species in the $n_{b4}/n_{b2}$ mixture
at a volume fraction of $\phi=0.5$. 
For all species, the repulsive force increases and reaches a peak (inflection point of the MRP) 
at distances less than $R_{gb\alpha}$, and then decays gradually, resulting in a tail.
The attractive force only appears at distances larger than $R_{gb\alpha}$.
It is noticeable that in such mixtures the peak of the force for the higher resolution species
appears at shorter distances, suggesting that the repulsive force for the higher level of CG is softer.
Figure \ref{nb4nb2} also compares the RDFs obtained from the IECG simulations in the canonical ensemble
with the ones obtained by the IECG theory for polymer mixtures discussed here.
First, excellent agreement is observed between the IECG simulation results and the IECG theoretical results.
More importantly, the difference between the atomistic simulation data and
the IECG results for all species lies within the statistical uncertainties. 
Note that the coordinates of the positions of the CG sites in atomistic simulations of polymer melts
were obtained by calculating the related centers of masses and in the aforementioned mixtures we assumed that 
the polymer chains in the atomistic simulations were randomly partitioned in a mixture of chains with four CG sites 
and chains with two CG sites, at the given volume fraction, $\phi$.  
Therefore, the MRPs used in the IECG mixture simulations 
indeed produce accurate RDFs consistent with the atomistic RDFs.

\begin{table}[tbh]
\centering
\caption{The calculated average pressure values for mixtures 
consisting of soft sphere/multiblob models with various resolutions
representing a polymer melt with $N=300$. 
All mixtures contain four-blob models ($n_{b4}$): $n_{b4}/n_{b1}$, $n_{b4}/n_{b2}$, $n_{b4}/n_{b6}$, and
$n_{b4}/n_{b10}$, where $n_{bx}$ indicates that the polymers are represented by $x$ number of blobs.
In particular, the pressure consistency is shown for various compositions for $n_{b4}/n_{b2}$ at 
$503$ K and $563$ K. Note that the atomistic pressure values for this polymer 
melt at $503$ K and $563$ K are $343\pm4$ and $630\pm3$ atm, respectively.
}
\begin{tabular}{ p{12.5mm}|p{12.5mm} p{12.5mm} p{12.5mm}  p{10mm} p{10mm} }
     \hline
     \multicolumn{6}{c}{ $n_{b4}/n_{b\beta}$ mixtures; $\phi_\alpha=0.5$; $T=503$ K }     \\
     \hline
     $n_{b\beta}$ & $1$ &  $2$  &  $6$ &  $10$ &  \\
     \hline
     $P_{\rm{theory}}$  & $343.5$ & $342.9$  & $341.8$ & $342.4$ &  \\
     $P_{\rm{sim}}$     & $343.7$ & $343.3$  & $343.4$ & $344.6$ &  \\
     \hline
     \multicolumn{6}{c}{ $n_{b4}/n_{b2}$ mixtures at various $\phi_\alpha$; $T=503$ K }     \\
     \hline
     $\phi_{\alpha}$ & $0$ &  $0.25$  &  $0.5$ &  $0.75$  & $1.0$  \\
     \hline
     $P_{\rm{theory}}$  & $343.3$ & $343.1$  & $342.9$ & $342.6$ & $342.0$ \\
     $P_{\rm{sim}}$     & $343.5$ & $343.3$  & $343.3$ & $343.2$ & $343.2$ \\
     \hline
     \multicolumn{6}{c}{ $n_{b4}/n_{b2}$ mixtures at various $\phi_\alpha$; $T=563$ K }     \\
     \hline
     $\phi_{\alpha}$ & $0$ &  $0.25$  &  $0.5$ &  $0.75$  & $1.0$  \\
     \hline
     $P_{\rm{theory}}$  & $629.2$ & $629.0$  & $628.6$ & $628.0$ & $626.5$  \\
     $P_{\rm{sim}}$     & $629.5$ & $629.3$  & $629.2$ & $629.2$ & $629.8$  \\
     \hline
\end{tabular}
\label{Pblob}
\end{table}

Furthermore, the pressure values can be calculated either from the IECG simulations and/or the IECG theory 
and compared with the atomistic simulations. The IECG theory starts from the statistical mechanical definition
of the pressure in the canonical ensemble, i.e. $P=k_{\rm{B}}T \left( \partial \ln Z/\partial V  \right)_{n,T}$,
where $Z$ is the classical configurational integral,\cite{Frenkel2002,Stillinger2002} and calculates 
the kinetic, intramolecular, and intermolecular contributions of the pressure making use of the intramolecular 
distribution functions that are obtained from the Gaussian statistics. 
Moreover, the IECG simulations use the virial equation to obtain the pressure. 
The top part of Table \ref{Pblob} reports 
the total pressure values for mixtures consisting of the $n_{b4}$ species
with a mole fraction of $0.5$ and other species with different granularities. As the level of CG decreases, the kinetic contributions increases and is compensated with the decrease in the virial pressure (see SM). 
Therefore, the IECG formalism results in values of the pressure that are,  
within a small margin of error, constant as the resolution of  one the species in the CG mixture is varied. In all cases the IECG pressure values  
lie within the statistical uncertainties of the ones obtained from atomistic simulations.

\begin{figure}[htb]
\includegraphics[width=\columnwidth]{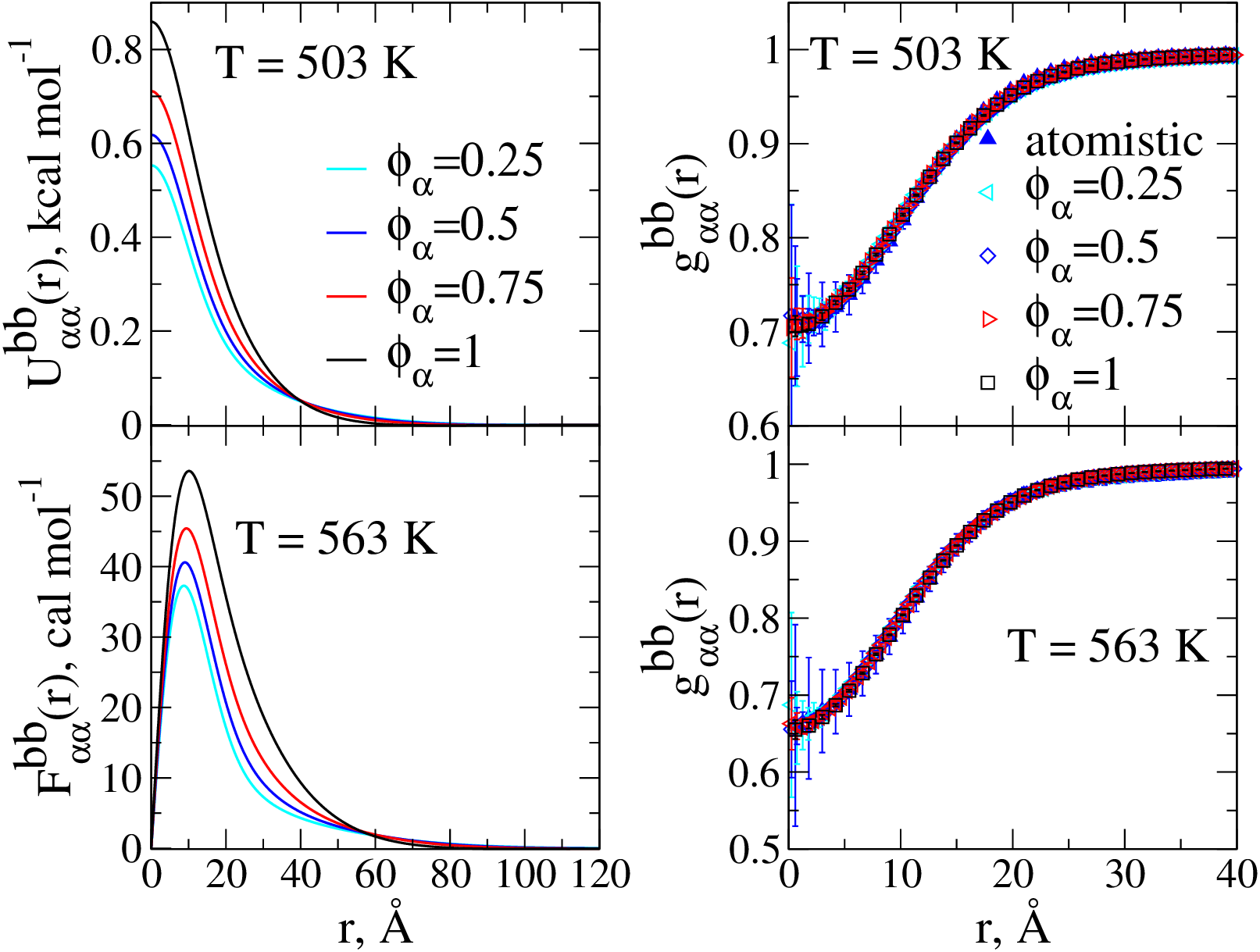}
\caption{ The radial distribution functions generated from composition dependent effective CG
potentials. Left panel: the numerical effective CG potentials for various compositions of the polymer
melts represented by di- and four blob models. The composition is shown in terms of the volume fraction for
four blob polymer melts,  $\phi_\alpha$.
Right panel: the radial distribution functions obtained from the IECG simulations using the effective CG potentials
for various compositions. Note that the atomistic radial distribution function is reported for the four blob model
corresponding to a volume fraction of one.
}
\label{nb4comp}
\end{figure}

It is important to show that the MRP at various compositions in multi-resolution simulations produces not only accurate pressure but also accurate structural properties.
The left panel of Figure \ref{nb4comp} shows the numerical MRP between the $n_{b4}$ species at two temperatures for
various compositions in a polymer mixture consisting of chains with two blobs and chains with four blobs.
As the volume fraction of high resolution species ($n_{b4}$) in the mixture increases,
the many-body effects of the $n_{b2}$ component become less dominant, decreasing the range of the potential,
while the values of the MRP at contact increase.
The right panel of Figure \ref{nb4comp} shows that the RDFs obtained
from the IECG simulations at various compositions at two temperatures 
are all in excellent 
agreement and they are within the uncertainties of atomistic simulations.
In the second and bottom panels of Table \ref{Pblob} the pressure values for a 
mixture of $n_{b4}$ and $n_{b2}$ species at variable volume fraction $\phi$ 
are presented at two different temperatures: 
all of the pressure values are within the statistical error of the related atomistic simulations.

The most challenging part in most multi-resolution simulations is the treatment of the mixed region, which is physically connecting simulations with different resolutions. When put into contact,  regions with different CG representations usually display different structural and thermodynamic properties, which lead to the emergence of unphysical forces in the mixed region.\cite{Potestio2013,DelleSite2007,Peters2016,Fritsch2012,Ensing2007}
In this work, we use liquid state theory principles to develop a multi-resolution description of a region where macromolecules are represented with CG models with different resolutions. Note that the properties of the mixed region are quantitatively consistent with the single-resolution CG samples and with the corresponding atomistic representations.
This consistency in pressure, radial distribution function, and molecular density allows liquids at various resolutions to be compatible with each other, opening up the possibility for future applications of this method in the contest of conventional multi-resolution methods.
We note that while in conventional multi-resolution simulations, atomistic and fine-grained resolutions are considered, here more extreme levels of coarse-graining are used and directly compared to the atomistic simulations. Selecting this model has the advantage of allowing for a formal, analytical solution to the problem.
We show that the effective CG potentials for the pure CG simulations are different from the MRPs for 
the CG multi-resolution simulations. 
Multi-resolution simulations that use the composition-dependent MRPs are shown 
to reproduce accurately the structural and thermodynamical properties of the atomistic 
polymeric liquid as well as of the pure CG simulations (see Figure \ref{nb4comp}) 
considering the relatively short-range repulsive 
and relatively long-range weak attractive CG interactions. 
Consequently, we show that significantly different MRPs generate almost 
indistinguishable pair correlations and pressure values. 
We note that the effective $\chi_{\rm{eff}}$ parameter\cite{Schweizer1997} defined by 
$\chi_{\rm{eff}} \propto c_{0\alpha\alpha} + c_{0\beta\beta} - 2 c_{0\alpha\beta}$ is zero for such systems, 
which suggests that the phase 
separation of species does not occur for such multi-resolution simulations as expected. 
Standard molecular dynamics algorithms can be used to integrate out the equation of motions for multi-resolution systems, 
but the timestep is restricted to the fastest motions of the highest resolution component. Multiple timestep 
methods such as the Reversible Reference System Propagator Algorithms\cite{Tuckerman1992} may be used to enhance computational 
efficiency.\cite{Dinpajooh2018} 
Therefore, the current formalism opens an avenue to rigorously use molecular simulations to 
study polymeric liquids at various resolutions in the length- and time scales relevant to advanced problems in polymer physics.

\vspace*{1.5cm}
\begin{acknowledgments}
This work was supported by the National Science Foundation (NSF) Grant No. CHE-1665466.
This work used the Extreme Science and Engineering Discovery Environment (XSEDE),\cite{xsede}
which is supported by the National Science Foundation Grant No. ACI-1548562.
This work used the XSEDE COMET at the San Diego Supercomputer Center through allocation
TG-CHE100082.
We are grateful to Paula J. Seeger for reading/editing the manuscript.
\end{acknowledgments}

\setcitestyle{square}
\bibliographystyle{apsrev4-1}

\end{document}